\documentclass[%
 reprint,
 amsmath,amssymb,
 aps,
]{revtex4-1}

\usepackage{graphicx}
\usepackage{dcolumn}
\usepackage{bm}
\usepackage[utf8]{inputenc}
\usepackage[spanish,english]{babel}
\usepackage[procnames]{listings}
\usepackage[svgnames]{xcolor}
\definecolor{lbcolor}{rgb}{0.9,0.9,0.9}
\usepackage{array}
\usepackage{enumitem}
\lstdefinestyle{Python} {
  language=Python,
  backgroundcolor=\color{lbcolor},
  basicstyle=\scriptsize\upshape\ttfamily,
  commentstyle=\color{blue},
  classoffset=1,
  morekeywords={Cosmology, pyCosmologyCBL},
  keywordstyle=\color{ForestGreen},
  classoffset=2,
  morekeywords={D_C},
  keywordstyle=\color{red},
  classoffset=0,
  tabsize=2,
  captionpos=b,
  frame=lines,
  numbers=left,
  numberstyle=\tiny,
  numbersep=5pt,
  breaklines=true,
  showstringspaces=false,
  procnamekeys={def,class}
}

\usepackage[none]{hyphenat}
\begin{document}
\preprint{APS/123-QED}

\title{Simulation of charged particles in Earth's magnetosphere: an approach to the Van Allen belts}

\author{J.E. García-Farieta}
\email{joegarciafa@unal.edu.co}%
\affiliation{%
Universidad Nacional de Colombia - sede Bogotá, Facultad de Ciencias, Departamento de física\\Av. Cra 30 No 45-03, Bogot\'a, Colombia
}
\author{A. Hurtado}%
\email{ahurtado@udistrital.edu.co}
\affiliation{%
Universidad Distrital Francisco José de Caldas, Facultad de Ciencias y Educación\\Proyecto Curricular de Licenciatura en física\\Grupo de Investigación FISINFOR\\Carrera 3 No 26A-40, Bogot\'a, Colombia
}

\date{\today}

\begin{abstract}{\centering{\bf Abstract}\\}
Earth's magnetosphere, beyond protecting the ozone layer, is a natural phenomena which allows to study the interaction between charged particles from solar activity and electromagnetic fields. In this paper we studied trajectories of charged particles interacting with a constant dipole magnetic field as first approach of the Earth's magnetosphere using different initial conditions. As a result of this interaction there is a formation of well defined radiation regions by a confinement of charged particles around the lines of the magnetic field. These regions, called Van Allen radiation belts, are described by classical electrodynamics and appear naturally in the numerical modeling done in this work.

\begin{description}
\item[PACS numbers] 02.60.Cb, 03.50.De, 41.20.Gz, 94.30.Va, 94.30.Xy, 94.30.Hn
\item[Keywords] Van Allen radiation belt, charged particles, Earth's magnetosphere, numerical modeling. 
\end{description}
\end{abstract}

\maketitle

\section{Introduction}

The interaction between charged particles and magnetic fields is one of the most interesting topics studied at undergraduate level in electromagnetism courses from a conceptual, experimental and computational point of view. This phenomenon opened the way to the modern physics as in the understanding of the special theory of relativity and quantum mechanics, it also supports a large number of experimental results and applications that can be seen in accelerators for high-energy physics experiments. Nevertheless the description on trajectories of charged particles presented in first courses of physics at university levels, is usually limited to the case of a uniform magnetic field, despite the fact that many textbooks described in previous sections systems with magnetic fields not uniform such is the case of the magnetic field produced by a current loop and even the dipole limit.\\

In this sense Earth is a perfect scenario to illustrate the trajectories of charged particles emitted by solar activity in interaction with Earth's magnetic field (called geomagnetic field). The behavior of charged particles in presence of the geomagnetic field has been widely studied \citep{dragt1965trapped}, coming to be a fundamental tool to understand the structure of the magnetosphere and the underlying phenomena to plasma physics generated by these types of interactions \citep{howard1999global, inarrea2004keplerian}. The Earth's magnetic field can be understood properly from the magnetohydrodynamics, in particular from the geodynamo theory, considering the Earth's core as a conducting fluid in rotation and with convective movements. This mechanism allows to describe at first approximation the geomagnetic field as the field produced by a magnetic dipole tilted 11.5º with respect to the Earth rotation axis and with a magnetic dipole moment $|\vec{m}|=7.79\times10^{22}$ Am${}^2$. Since this field changes slowly over the years, it produces a secular drift of the magnetic poles and there is consequently a decrease of the magnetic dipole moment \cite{olson2006changes}. In this work these effects have been neglected for pedagogical purposes to have a constant magnetic field.\\

The movement of charged particles trapped in the geomagnetic field has been a study interest for physics, Earth sciences and engineering because of its relationship with the phenomenon of auroras, cosmic rays and radiation belts. A first mathematical formulation of this problem, considering a dipole magnetic field, was established by St\"oermer \cite{stormer1907trajectoires, stormer1934trajectoires}. As is well known, in general the solution of the equations of motion to describe the trajectory of a charged particle in a magnetic field is not always easy to obtain and in many cases you can only solve the problem using the numerical integration of these equations.\\

\begin{figure}[h!t]
\centering
\includegraphics[width=\linewidth]{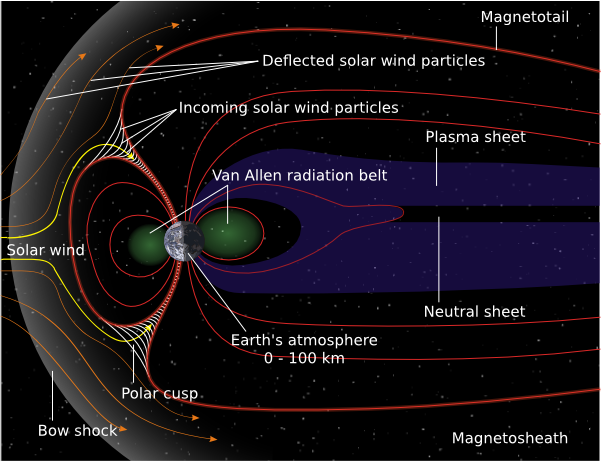}
\caption{Earth's magnetosphere. Interaction with solar wind and Van Allen belts can be appreciated. Figure obtained from \cite{NASA}. }\label{fig:magnetosfera}
\end{figure}

A particular example illustrated in introductory textbooks is the interaction of a charged particle with a static and uniform magnetic field, in which case the trajectory can be obtained analytically since the equations of motion can be easily uncoupled, such, for example, circular motion or helices around the magnetic field lines. In a non-uniform magnetic field the motion can be more complex to describe, even under certain conditions the particles can be trapped around the magnetic field lines forming stable confinement regions, this is what happens for instance with part of the solar wind interaction with the Earth's magnetic field and which originate the so-called Van Allen belts. The Fig. \ref{fig:magnetosfera} shows schematically different regions of the Earth's magnetosphere, Van Allen belts (in green) form toroidal rings mainly composed by protons, electrons and $\alpha$ particles. Two belts can be distinguished, an internal one that extends from 0.2 to 2 Earth radii ($R_E$), and an external one from 3 to 10 Earth radii.\\

In accordance with the above and with pedagogical purposes, the motion of a charged particle interacting with a magnetic dipole is an illustrative and conceptually rich example that allows to take advantage of computational tools to solve a problem with no analytical solution. Moreover it provides several advantages to support lessons allowing manipulating different parameters to define the physical system and characterize the particle trajectories. In this paper a non-relativistic version has been considered, i.e. radiation emitted by accelerated particles is not taken into account in order to simplify the model and make it suitable to courses on electromagnetic theory. In the next sections the theoretical model is developed introducing the concept of magnetic moment and a numerical implementation is carried out to finally discuss the obtained results.

\section{Theoretical description}
\begin{figure}[h!t]
\centering
\includegraphics[width=72mm]{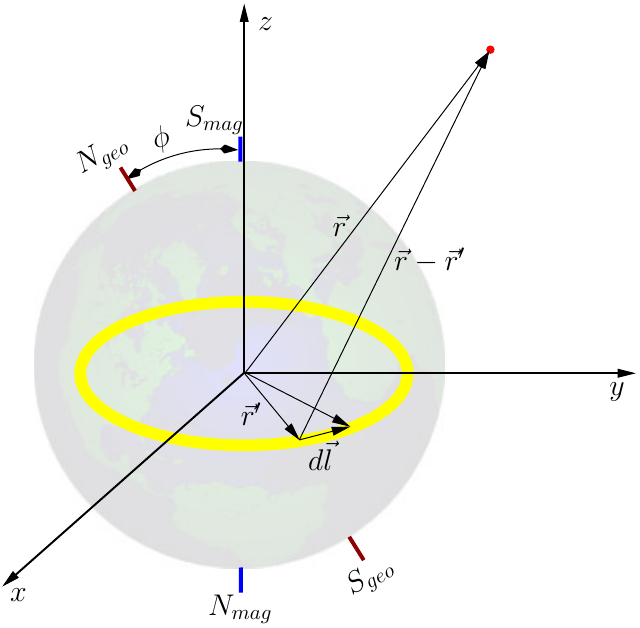}
\caption{Dipole model of the Earth's magnetic field from a current loop in approximation for far points ($r'/r\ll1$).}\label{fig:EarthsRing}
\end{figure}

In order to model the dipole magnetic field mentioned above, a circular loop with arbitrary current is assumed in the Earth's core with a certain angle of inclination as Fig. \ref{fig:EarthsRing} shows. In this approach the geometry and topographic deformations of the Earth do not affect the source of magnetic field, due to this the particle trajectories is determined only by the interaction with the magnetic field neglecting any effect by gravitational interactions, self-interactions between particles or external fields. The magnetic field produced by a magnetic dipole is well known \citep{Davis2009}, in its coordinate-free form is given by the equation (\ref{eq:Bdipole}).

\begin{equation}\label{eq:Bdipole}
\vec{B}=\frac{\mu_0}{4\pi r^3}[3(\vec{m}\cdot\hat{r})\hat{r}-\vec{m}]
\end{equation}

where $\mu_0$ is the permeability of free space, $\vec{m}$ the magnetic dipole moment, $\vec{r}'$ and $\vec{r}$ vectors measured from the origin of the reference system to an element of the current loop and to an arbitrary point in space respectively. The dipole limit is obtained under the assumption $r'/r\ll1$, as consequence $\vec{m}$ will be aligned along the the loop's symmetry axis. From a pedagogical point of view equation (\ref{eq:Bdipole}) can be a little confusing for students.
A key problem with is that usually introductory textbooks on electromagnetic theory only consider magnetic fields at symmetry points of the systems, moreover many of them introduce superficially the concept of magnetic dipole moment as a result after calculating the torque of a current loop in an uniform magnetic field. An even greater source of concern is that this equation, which is fundamental in magnetostatics, does not even appear in many general physics textbooks.\\

The magnetic field due to dipole can be obtained by several ways, all of these descriptions are equivalent between them but with different level of complexity. Some of these versions are:

\begin{enumerate}[label={\bf \Alph*)}]
\item Using the Biot-Savart law [equation (\ref{eq:BiotSavart})] to compute the magnetic field in an point outside of the current loop axis and then using the far-field approximation for points located at $r\gg r'$ (dipole limit). In this case elliptic integrals will appear naturally and even if they are solved using approximations, the procedure in general 
is extremely time-consuming given the math details.
\item It is possible to compute the vector potential $\vec{A}$ for a magnetic dipole moment $\vec{m}$ and then to write the magnetic field induction as $\vec{B}=\nabla\times\vec{A}$. Although this approach is interesting and mathematically simple, it implies having previously introduced the concepts of $\vec{A}$ y $\vec{m}$, it is rarely done in introductory textbooks.
\item Alternatively is possible to get $\vec{B}$ from the magnetic scalar potential $\Phi$ considering that space has not free currents. Consequently the intensity of magnetic field $\vec{H}$ satisfies $\nabla\cdot\vec{H}=\mu_0^{-1}\nabla\cdot\vec{B}=0$ and also Laplace equation, so that $\vec{B}=\mu_0^{-1}\nabla\Phi$. Although $\Phi$ can be easily obtained by solving the Laplace's equation with appropriate boundary conditions and considering the condition of continuity of the magnetic field, this development is more suitable for later courses on electromagnetism and classical electrodynamics theory.
\end{enumerate}

\begin{equation}\label{eq:BiotSavart}
\vec{B}(\vec{r})=\frac{\mu_0I}{4\pi}\oint_C\frac{d\vec{l}\times(\vec{r}-\vec{r}')}{|\vec{r}-\vec{r}'|^3}
\end{equation}

In order to obtain the expression (\ref{eq:Bdipole}) in a simple way, without loss of generality, we have chosen to follow the main idea expressed in the case \emph{A)} but adopting the approach proposed by \cite{Bezerra2012}, this solution makes the mathematical procedure shorter and it is focused on physical concepts. Considering the Fig. \ref{fig:EarthsRing} taking into account distant points from the current ring $r'/r\ll1$, a Taylor expansion can be done directly in the denominator of the Biot-Savart law (\ref{eq:BiotSavart}). This leads to the following results and finally to the equation (\ref{eq:BdipoleDesarrollo}) which matches the desired result.

\begin{eqnarray}\label{eq:BdipoleDesarrollo}
\vec{B}(\vec{r})&\simeq&\frac{\mu_0I}{4\pi}\oint_C\frac{d\vec{l}\times(\vec{r}-\vec{r}')}{|\vec{r}-\vec{r}'|^3}\nonumber\\
&=&\frac{\mu_0I}{4\pi r^3}\oint_C d\vec{l}\times(\vec{r}-\vec{r}')\left(1+3\frac{\hat{r}\cdot\vec{r}'}{r}\right)\nonumber\\
&=&\frac{\mu_0I}{4\pi r^3}\left[-\oint_C d\vec{l}\times\vec{r}'+3\left(\oint_C d\vec{l}(\hat{r}\cdot\vec{r}')\right)\times\hat{r}\right]\nonumber\\
&=&\frac{\mu_0I}{4\pi r^3}\left\{\oint_C \vec{r}'\times d\vec{l}+3\left[\left(\frac12\oint_C \vec{r}'\times d\vec{l}\right)\times\hat{r}\right]\times\hat{r}\right\}\nonumber\\
&=&\frac{\mu_0}{4\pi r^3}[2\vec{m}+3(\vec{m}\times\hat{r})\times\hat{r}]\nonumber\\
&=&\frac{\mu_0}{4\pi r^3}\left[3(\vec{m}\cdot\hat{r})\hat{r}-\vec{m}\right],
\end{eqnarray}

where the magnetic moment has been naturally defined as $\vec{m}=\frac{I}{2}\oint_c\vec{r}'\times d\vec{l}$, turning out to be axial to the plane of current loop since $\vec{r}'$ and $d\vec{l}$ are always perpendicular, as can be verified by applying the right-hand rule. In addition, the double cross product was used $(\vec{m}\times\hat{r})\times\hat{r}=(\vec{m}\cdot\hat{r})\hat{r}-\vec{m}$ and the vector identity $\oint_c (\hat{r}\cdot\vec{r}')d\vec{l}=\left(\frac12\oint_c\vec{r}'\times d\vec{l}\right)\times\hat{r}$. 

\begin{figure}[h!t]
\centering
\includegraphics[width=\linewidth]{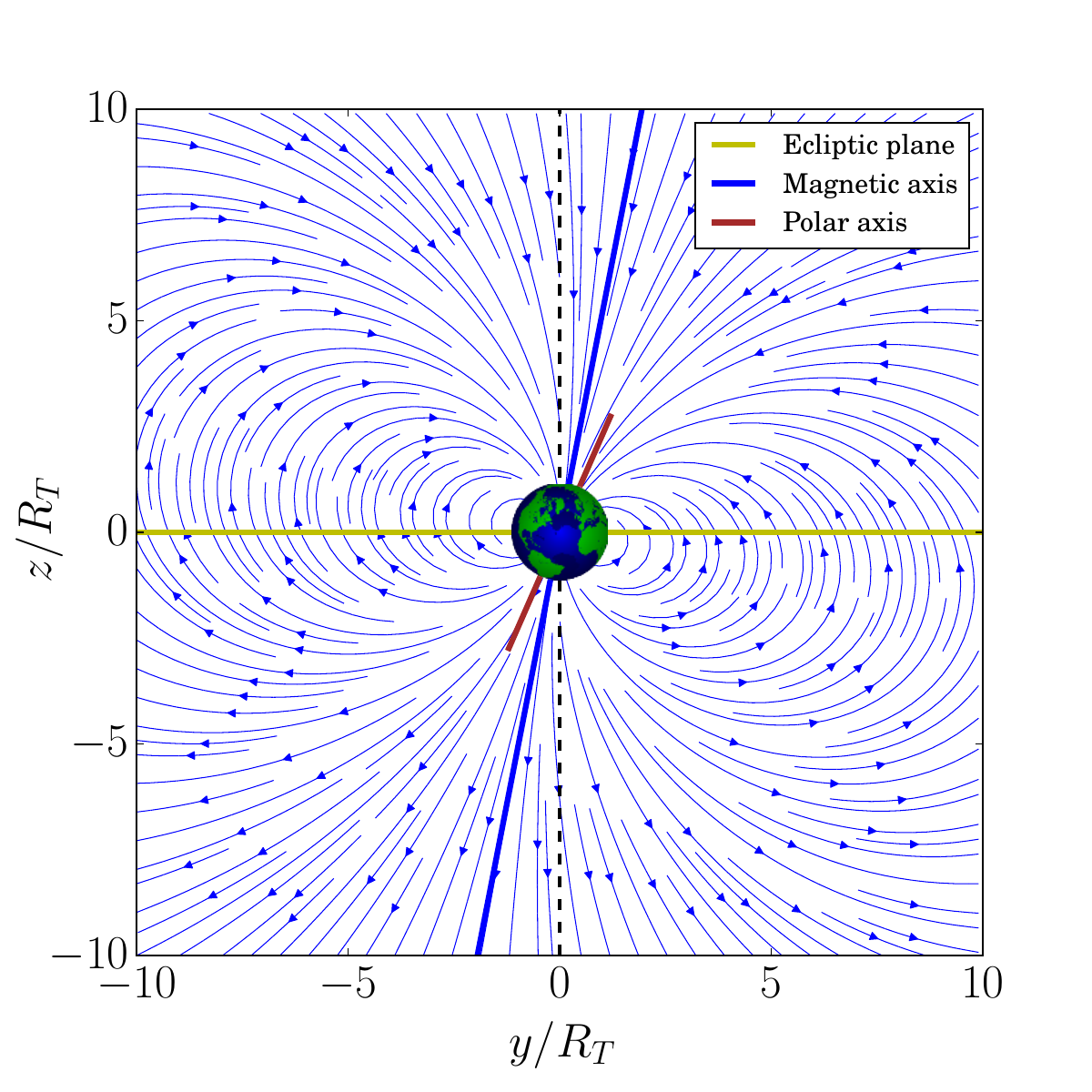}
\caption{Geomagnetic field lines in the dipole approximation.}\label{fig:lineascampo}
\end{figure}

\begin{figure}[h!t]
\centering
\includegraphics[width=\linewidth]{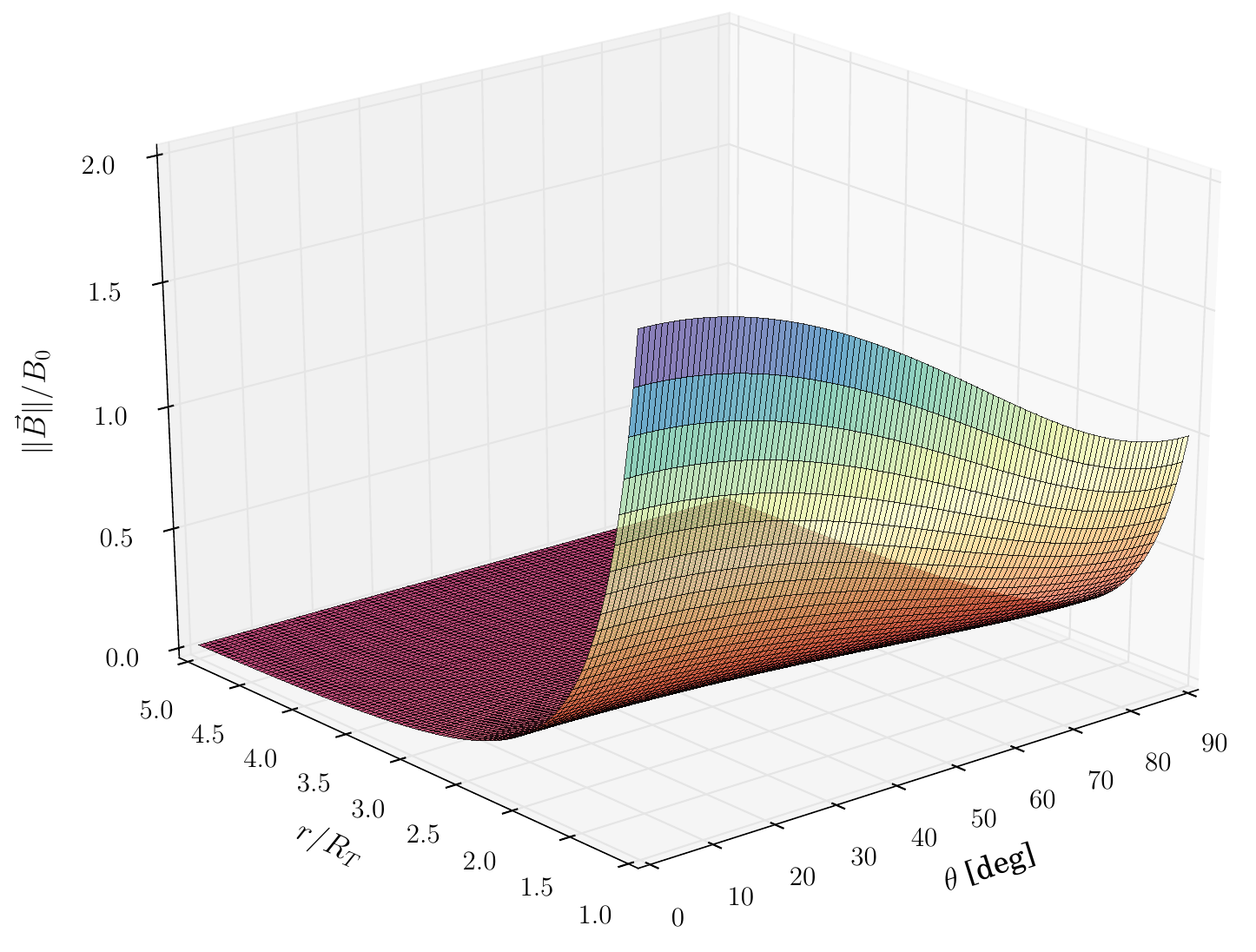}
\caption{Magnetic field magnitude of a dipole as a function of the radial distance $r$ (in Earth radius) and polar angle $\theta$, with $B$ measured in units of the equatorial field $B_0$.}\label{fig:MagnitudeBfield}
\end{figure}

The Earth's magnetic field lines in the dipole approximation (\ref{eq:Bdipole}) are shown in Fig. \ref{fig:lineascampo}. As seen in it the rotation axis is inclined $\theta=23.5^\circ$ with respect to the ecliptic plane (in yellow, this plane defines the orbit of the Earth around the Sun), and at the same time the magnetic poles are shifted $\phi=11.5^\circ$ from the geographical poles. The magnetic field intensity at a point is given by (\ref{eq:dipoleintensity}). This expression can be easily obtained from (\ref{eq:Bdipole}) considering the representation in spherical coordinates [equation (\ref{eq:dipoleesfericas})] or Cartesian coordinates [equation (\ref{eq:dipolecartesianas})], this last coordinate system was used to perform the simulation. Fig. \ref{fig:MagnitudeBfield} shows how the intensity of the Earth's magnetic field (normalized to the mean value of the equatorial magnetic field in the surface $B_0=3.12\times10^{-5}T$), is changing in function of coordinates $r$ (measured in Earth radius $R_T$) and $\theta$ (in degrees). In particular, from the set of equations (\ref{eq:dipoleesfericas}) it follows that the dipolar field is parallel to the radial direction on the poles and perpendicular to the radial direction at the equator, also, at a fixed distance the field is twice as intense at the poles than in the equator and the field strength decay with the radial distance as $1/r^3$. In the equation (\ref{eq:dipoleintensity}) is explicit the dependence between field intensity, distance and angle measured from the pole (latitude), this behavior is also illustrated in Fig. \ref{fig:MagnitudeBfield}.

\begin{table*}[!tbp]
  \centering
  \begin{tabular}{lcc}
    \begin{minipage}{.3\textwidth}
    \begin{eqnarray}
    r_0&=&(-4.,~-1.0,~-6.0)~[R_T]\nonumber\\ v_0&=&(0.1,~0.1,~0.1)~[m/s]\nonumber
    \end{eqnarray}
      \includegraphics[width=\linewidth]{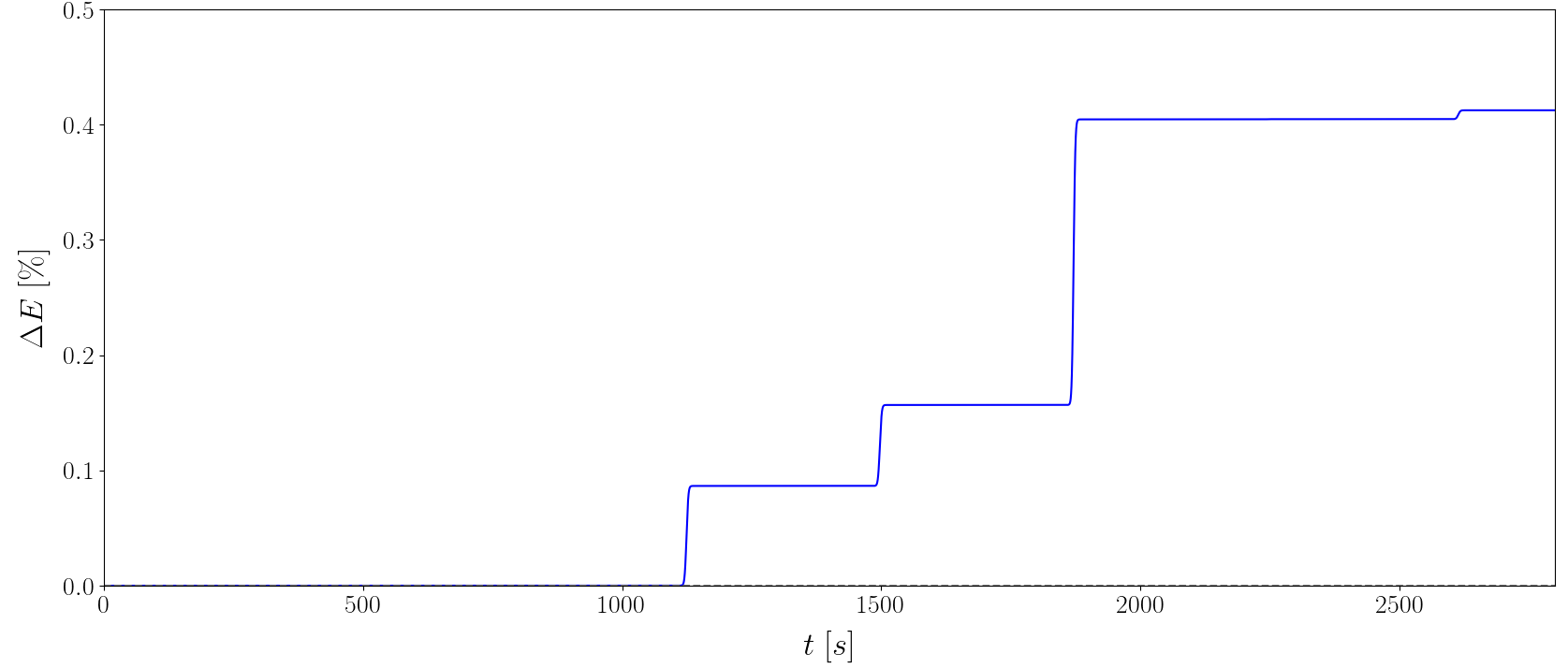}
    \end{minipage}
    &
    \begin{minipage}{.3\textwidth}
      \includegraphics[width=\linewidth, height=50mm]{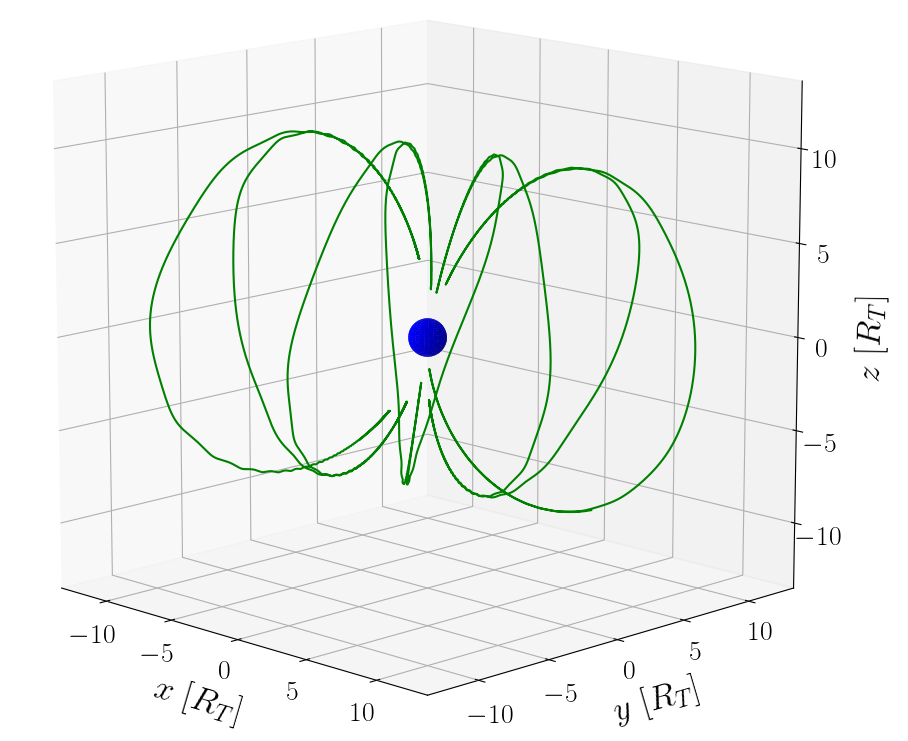}
    \end{minipage}
    & 
    \begin{minipage}{.3\textwidth}
      \includegraphics[width=\linewidth, height=50mm]{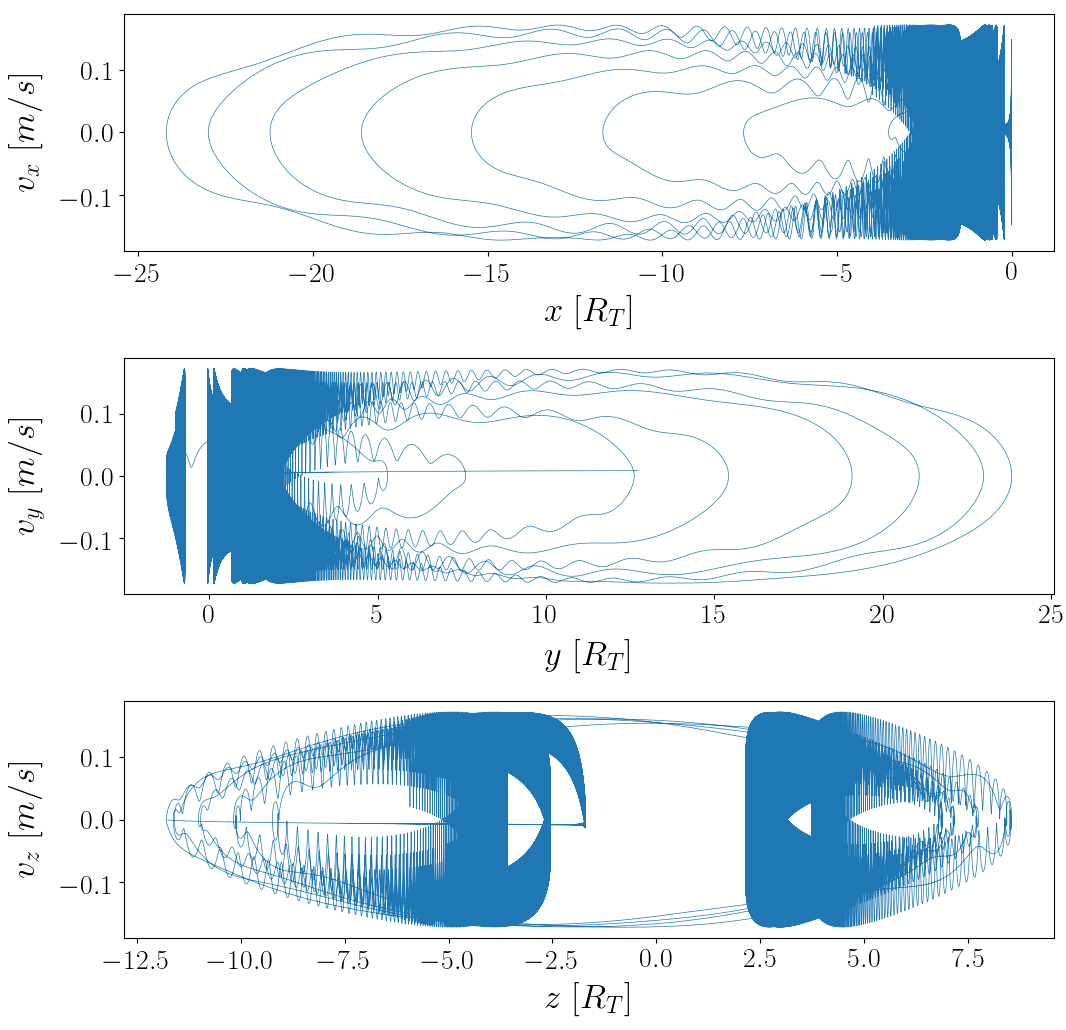}
    \end{minipage}
 \\
    \begin{minipage}{.3\textwidth}
    \begin{eqnarray}
    r_0&=&(0.0,~-7.85,~-1.53)~[R_T]\nonumber\\ v_0&=&(0.0,~0.3,~0.3)~[m/s]\nonumber
    \end{eqnarray}
      \includegraphics[width=\linewidth]{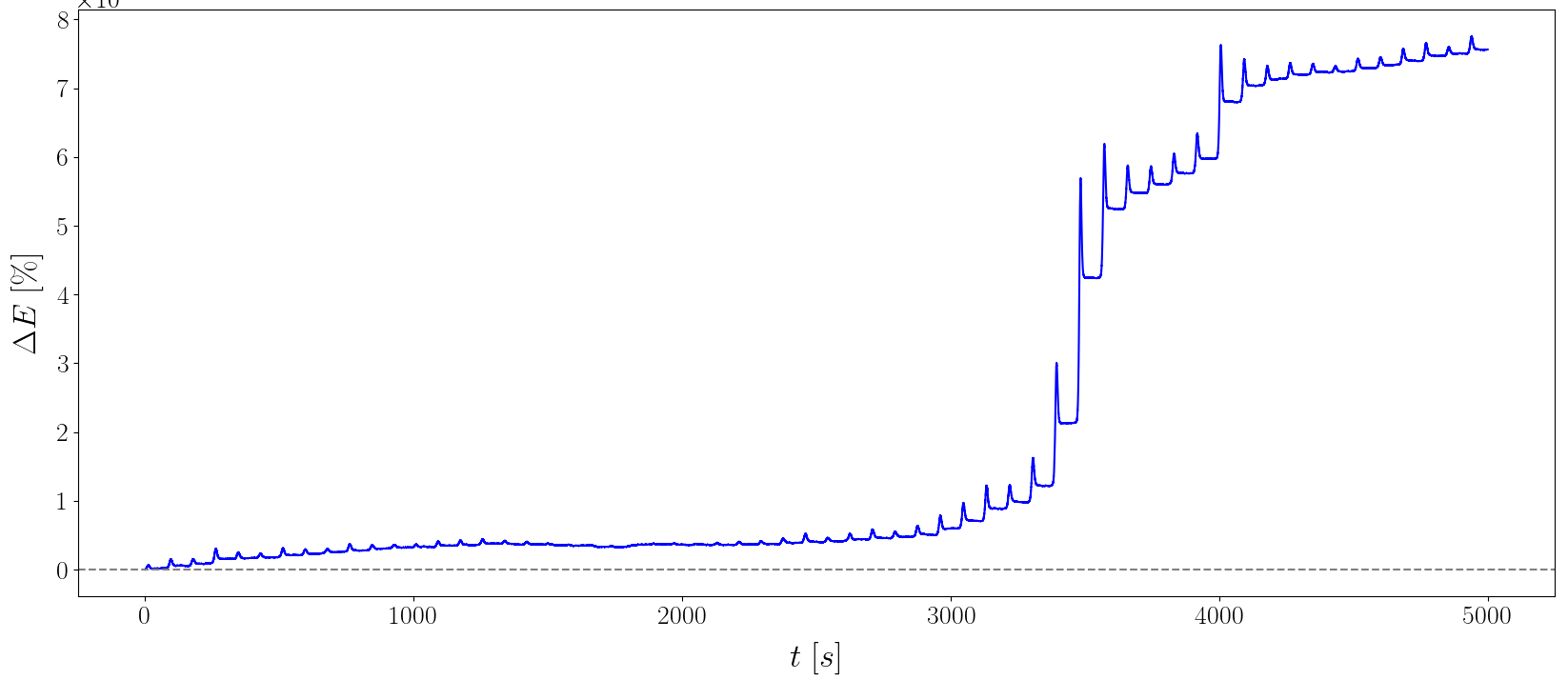}
    \end{minipage}
    &
    \begin{minipage}{.3\textwidth}
      \includegraphics[width=\linewidth, height=50mm]{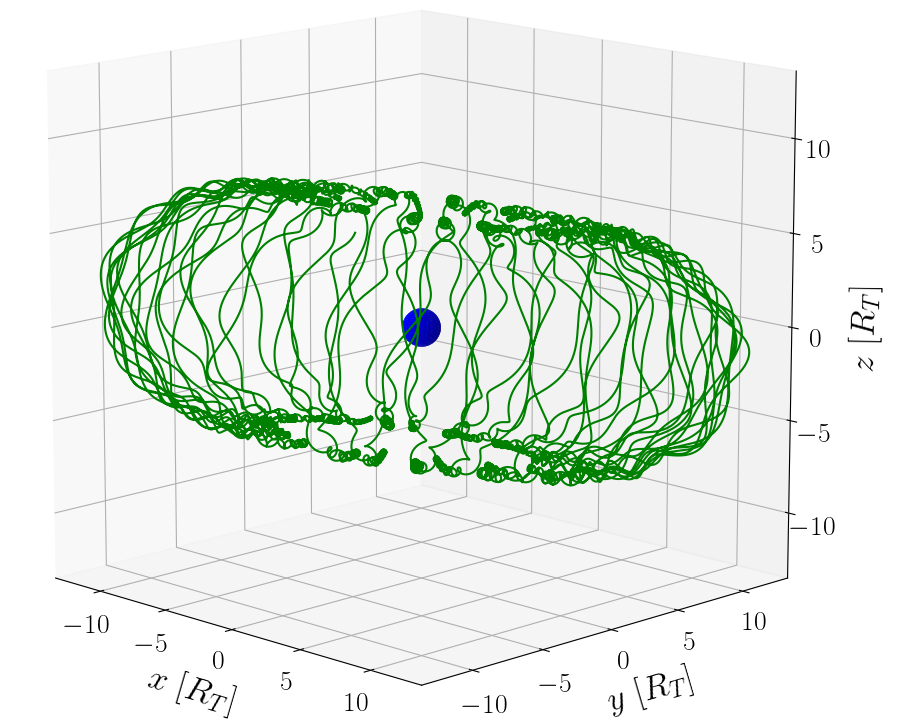}
    \end{minipage}
    & 
    \begin{minipage}{.3\textwidth}
      \includegraphics[width=\linewidth, height=50mm]{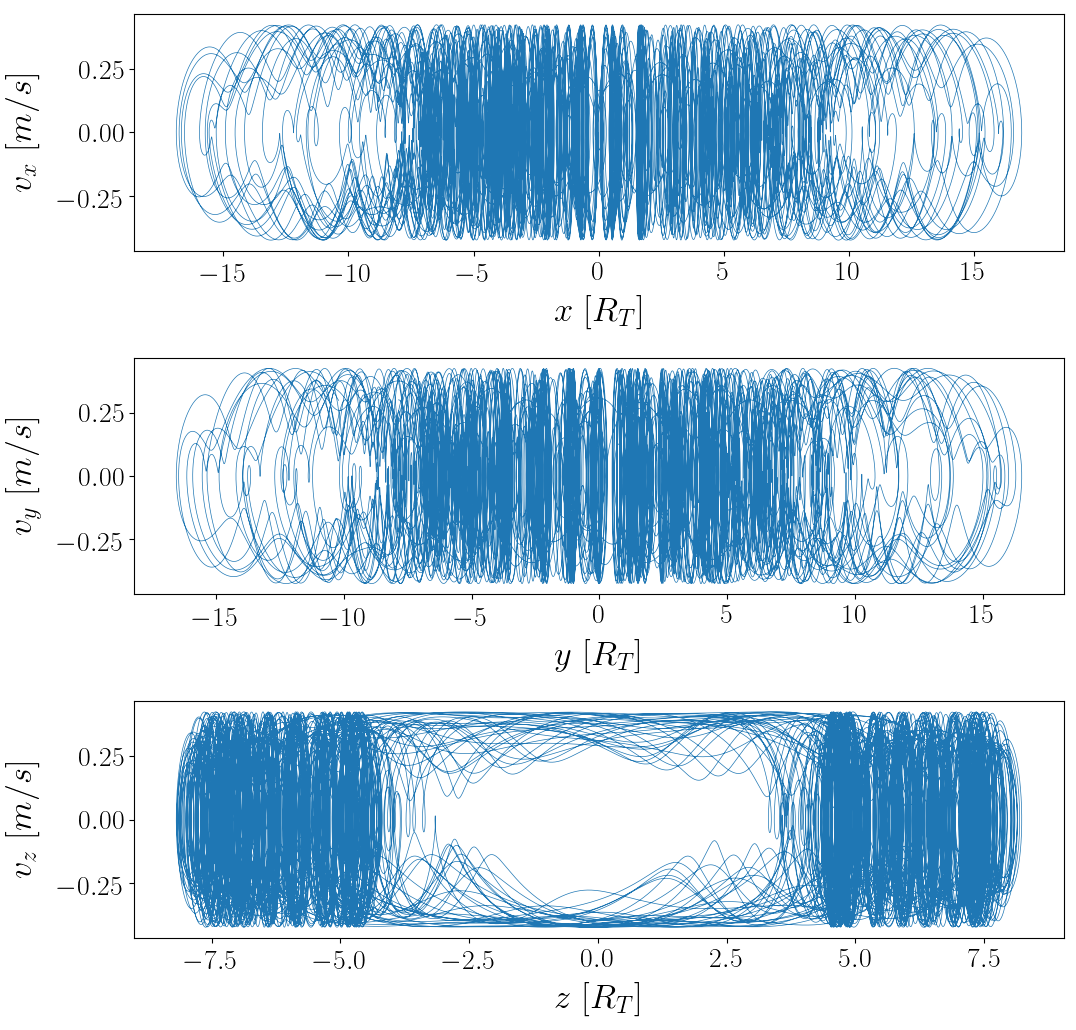}
    \end{minipage}
 \\
    \begin{minipage}{.3\textwidth}
        \begin{eqnarray}
    r_0&=&(-35.,~-30.0,~-35.0)~[R_T]\nonumber\\ v_0&=&(2.0,~1.0,~0.0)~[m/s]\nonumber
    \end{eqnarray}
      \includegraphics[width=\linewidth]{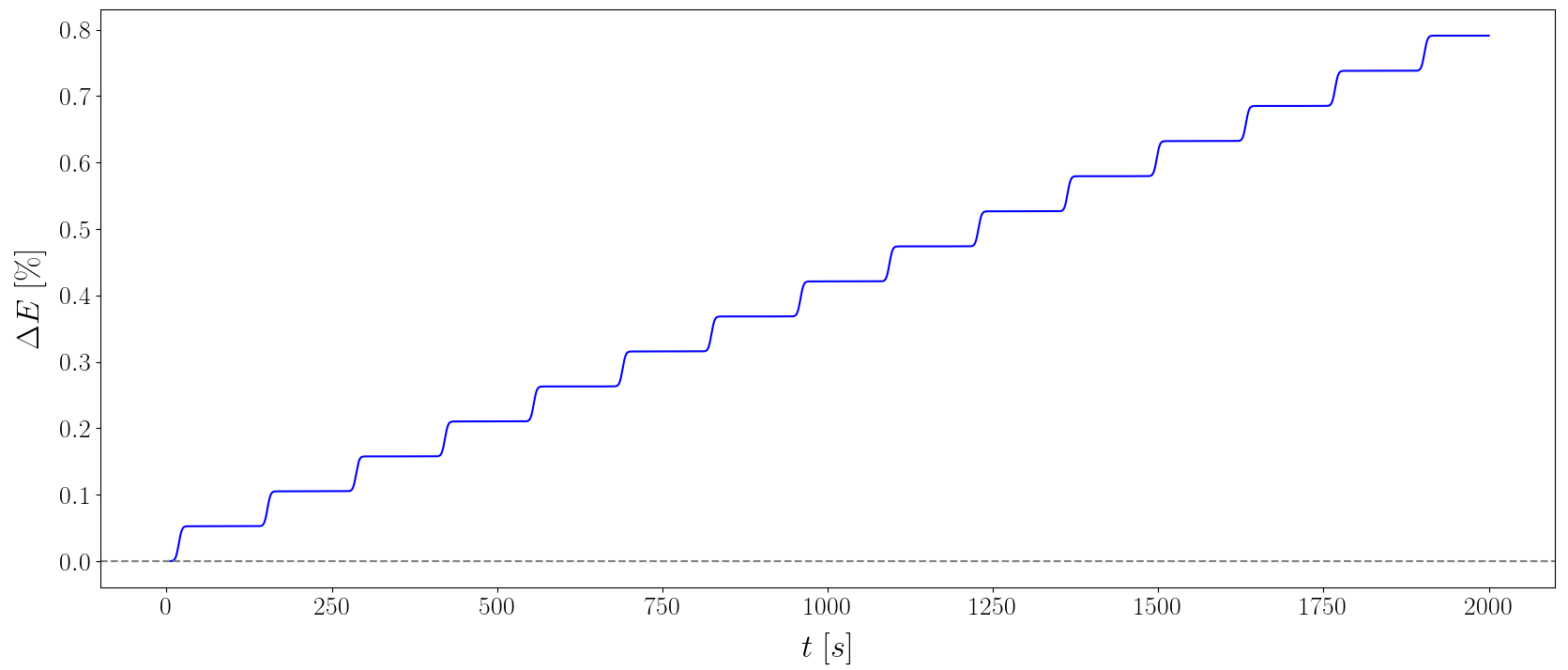}
    \end{minipage}
    &
    \begin{minipage}{.3\textwidth}
      \includegraphics[width=\linewidth, height=50mm]{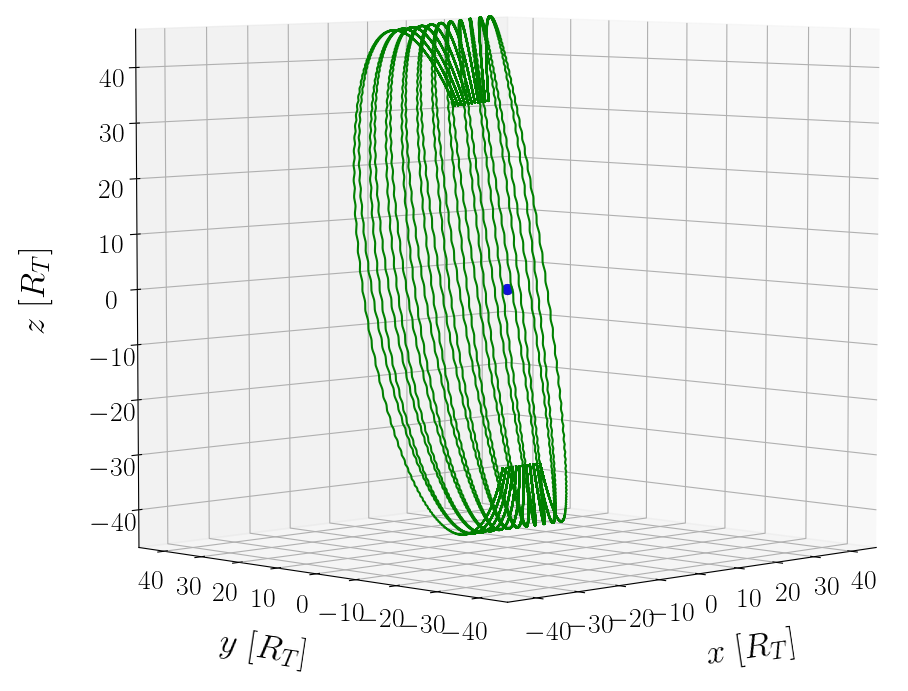}
    \end{minipage}
    & 
    \begin{minipage}{.3\textwidth}
      \includegraphics[width=\linewidth, height=50mm]{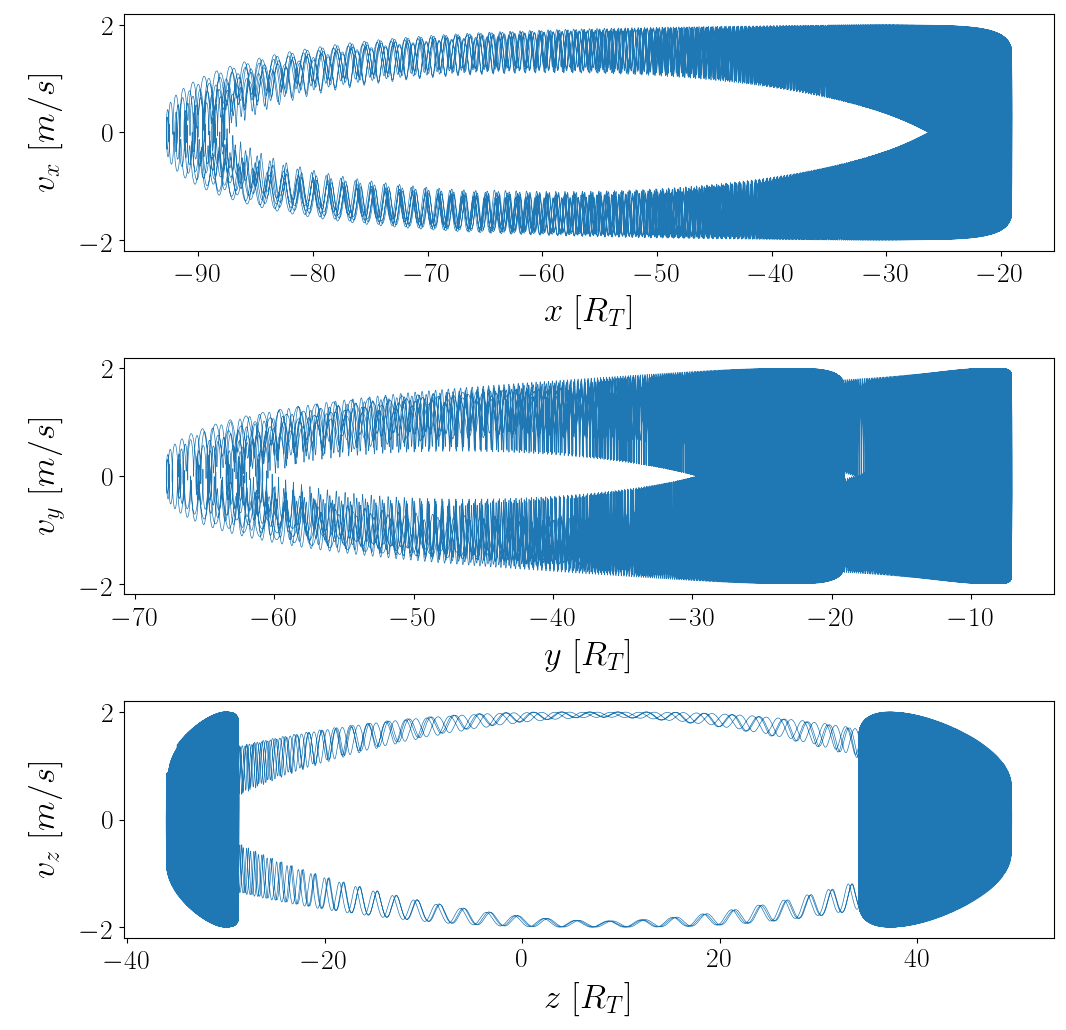}
    \end{minipage}
 \\
    \begin{minipage}{.3\textwidth}
        \begin{eqnarray}
    r_0&=&(-30.,~-30.0,~-30.0)~[R_T]\nonumber\\ v_0&=&(-2.0,~-1.0,~10.0)~[m/s]\nonumber
    \end{eqnarray}
      \includegraphics[width=\linewidth]{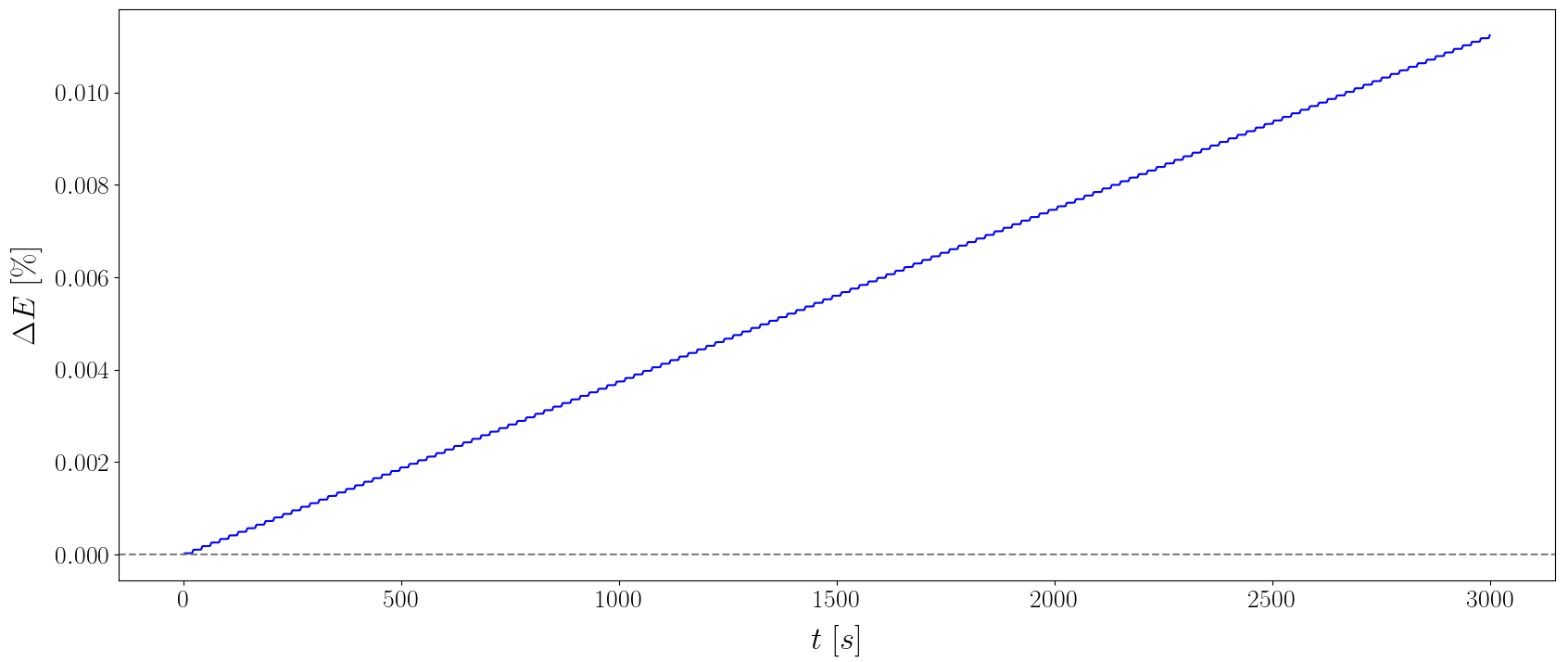}
    \end{minipage}
    &
    \begin{minipage}{.3\textwidth}
      \includegraphics[width=\linewidth, height=50mm]{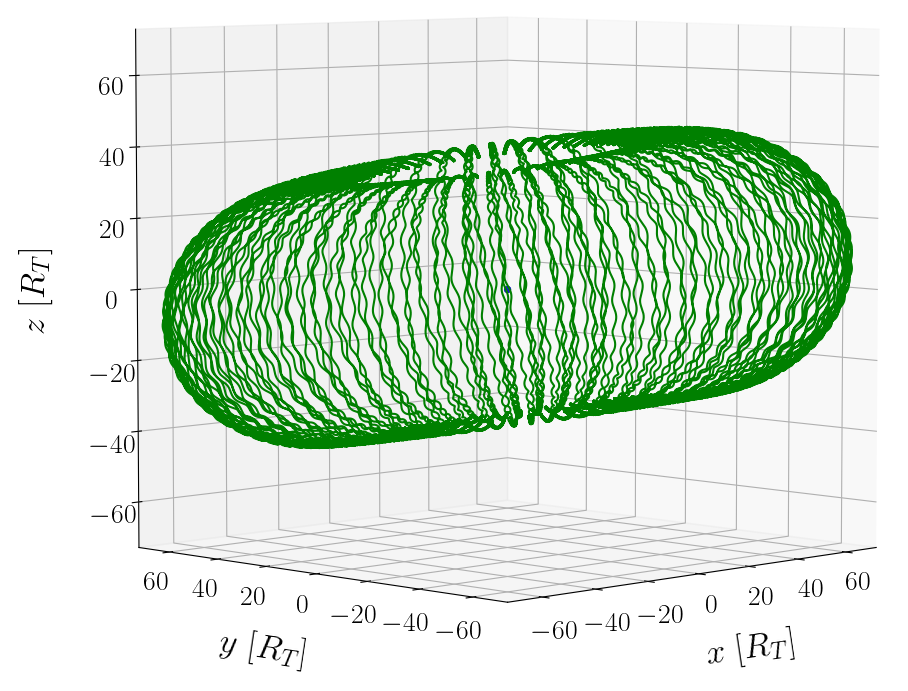}
    \end{minipage}
    & 
    \begin{minipage}{.3\textwidth}
      \includegraphics[width=\linewidth, height=50mm]{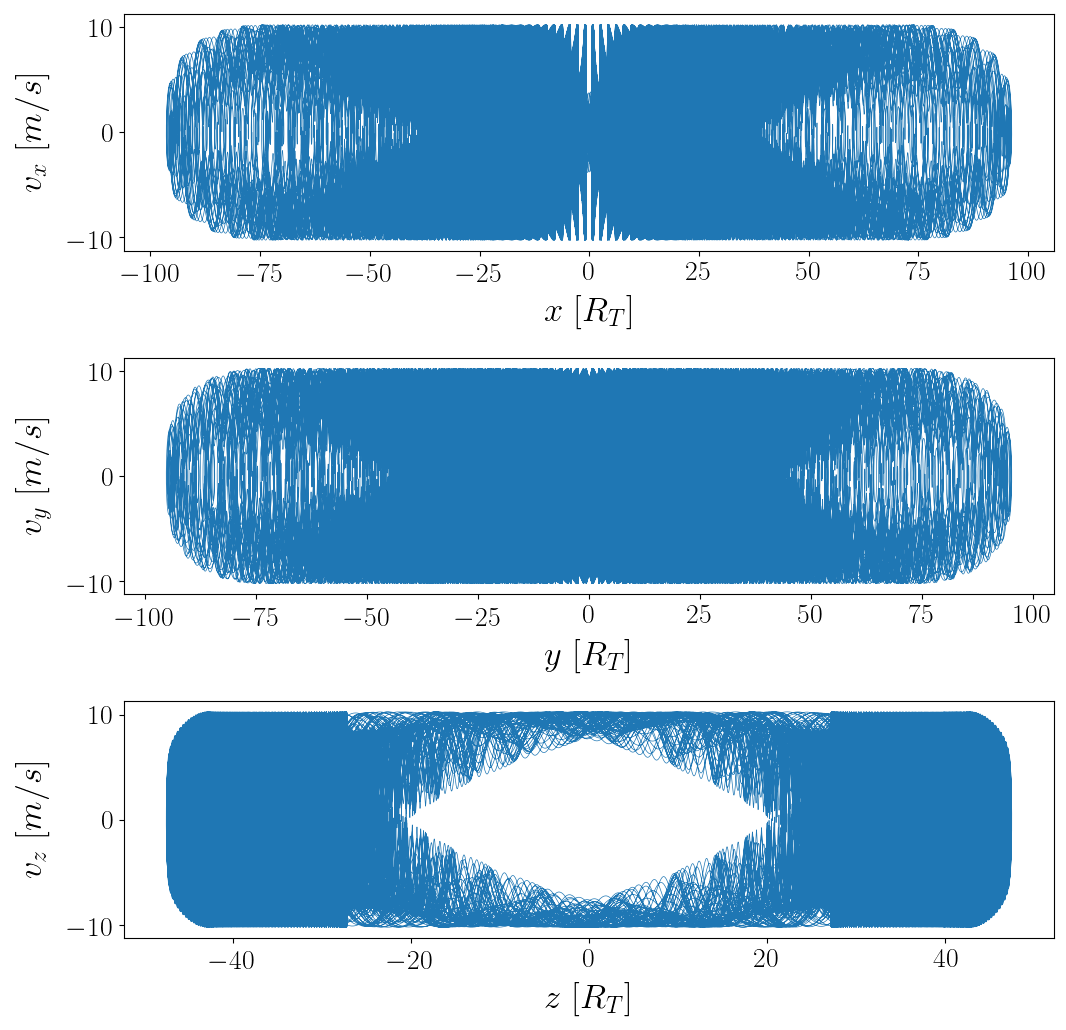}
    \end{minipage}
  \end{tabular}
  \caption{Initial conditions, particle trajectory, phase-space for each coordinate and total energy diagram as function of time}\label{Tab:TrajectoriesParticles}
\end{table*}

\begin{equation}\label{eq:dipoleintensity}
  \lVert{B}\rVert=\frac{m\mu_0}{4\pi r^3}(1+3\cos^2\theta)^{1/2}
\end{equation}

\noindent
\begin{tabular}{@{}p{.5\linewidth}@{}p{.5\linewidth}@{}}
\begin{subequations}\label{eq:dipolecartesianas}
\begin{eqnarray}
  B_x&=&\frac{3m\mu_0}{4\pi r^5}xz\\
  B_y&=&\frac{3m\mu_0}{4\pi r^5}yz\\
  B_z&=&\frac{m\mu_0}{4\pi r^5}(z^2-r^2)
\end{eqnarray}
\end{subequations}
  &
\begin{subequations}\label{eq:dipoleesfericas}
\begin{eqnarray}
  B_r&=&\frac{2m\mu_0}{4\pi r^3}\cos\theta\\
  B_\theta&=&\frac{m\mu_0}{4\pi r^3}\sin\theta\\
  B_\phi&=&0
\end{eqnarray}
\end{subequations}
\end{tabular}

The dynamics of the charged particles interacting with the magnetosphere is determined by the Lorentz force in the absence of electric fields $\vec{F}=q\vec{v}\times\vec{B}$. Neglecting relativistic effects, the equations of motion for a particle with charge $q$ and mass $m_p$ is (\ref{eq:Newlorentzforce}).

\begin{equation}\label{eq:Newlorentzforce}
\frac{d^2\vec{r}}{dt^2}=\frac{q}{m}\frac{\mu_0}{4\pi r^3}\frac{d\vec{r}}{dt}\times[3(\vec{m}\cdot\hat{r})\hat{r}-\vec{m}].
\end{equation}

To solve this differential equation one could think about rewriting it appropriately in spherical or cylindrical coordinates taking advantage of the angular symmetry of the dipole in $\theta$ and $\phi$, and the fact that $\vec{m}$ is constant and oriented on the magnetic axis. Even under this assumption, the equation (\ref{eq:Newlorentzforce}) represents a coupled system of non-linear first-order differential equations due to the shape of the magnetic field, which makes it non-analytically integrable \citep{saletan1998classical}.

\section{Simulation and numerical method}
The trajectories of interacting particles with the Earth's magnetosphere are given by the solution of equation (\ref{eq:Newlorentzforce}) for a flow of particles since solar wind is basically a plasma. The simplest scheme to describe this plasma is to consider that it is composed of independent particles neglecting any other interaction and surrounding effects (``\emph{Single particle motion}''). This is a very simplified version of the real case but it has the advantage of illustrating the types of trajectories formed depending on the initial conditions and the formation of confinement regions without entering into statistical descriptions, therefore it is good enough for purposes of the present work.\\

There is a vast amount of literature on integration schemes to solve numerically the equation (\ref{eq:Newlorentzforce}) \citep{betts1998survey, mazumdar1988analysis, mcguire2003using}. Furthermore recent research focuses on the optimization of several methods for trajectories of charged particles in magnetic fields \citep{Zhaoqi2017, ioanoviciu2015new}. As is well-known Runge-Kutta fourth-order method (RK4) is one of the fixed-step integrators most used in science and engineering to get accurate results in problems without too much complexity. This method has a significantly higher convergence for orbital problems compared with Euler method, its global error is $dt^4$ and its truncation error is $O(dt^5)$ being $dt$ the step-size of the iteration. RK4 is also easy to implement computationally in high or medium level languages such as Java, Fortran, C/C++, Matlab, Mathematica, Python among others. For the purposes of this paper the RK4 method has been implemented in Python to determine the trajectories of interest, being necessary to rewrite (\ref{eq:Newlorentzforce}) as a system of two first order differential equations, as shown in equations (\ref{eq:RK4_1}) and (\ref{eq:RK4_2}).

\begin{subequations}\label{eq:RK4}
\begin{eqnarray}
  \frac{d\vec{v}}{dt}&=&\frac{q}{m_p}\frac{\mu_0}{4\pi r^3}\frac{d\vec{r}}{dt}\times[3(\vec{m}\cdot\hat{r})\hat{r}-\vec{m}]\label{eq:RK4_1}\\
  \vec{v}&=&\frac{d\vec{r}}{dt}\label{eq:RK4_2}
\end{eqnarray}
\end{subequations}

Following the RK4 scheme, an ordinary differential equation of first-order $\frac{du}{dt}=f(t,u)$ with initial condition $u(t_0)=u_0$ can be expressed as equation (\ref{eq:RK4metodo1}),

\begin{equation}\label{eq:RK4metodo1}
u_{k+1}=u_k+\frac{dt}{6}(k_1+2k_2+2k_3+k_4).
\end{equation}

In such a way that the value $u_{k+1}$ depends on the pre-existing value $u_k$ plus a slope with coefficients $k_i$ given by (\ref{eq:RK4metodo2}). The slope can be understood as a weighted average of slopes: at the beginning of the interval $k_1$, in the middle of the interval ($k_2$, $k_3$) and at the end of the interval $k_4$.

\begin{equation}\label{eq:RK4metodo2}
  \begin{aligned}
k_1&=f(t_k,~u_k)\\
k_2&=f\left(t_k+\frac{dt}{2},~u_k+\frac{dt}{2}k_1\right)\\
k_3&=f\left(t_k+\frac{dt}{2},~u_k+\frac{dt}{2}k_2\right)\\
k_4&=f(t_k+dt,~u_k+dtk_3)
  \end{aligned}
\end{equation}

An important detail about the simulation is the normalization of constants and parameters. This procedure is usually done in order to: use the characteristic scales of the physical phenomenon under study, reduce the numerical complexity and to facilitate the visualization of results. In any case the equations of motion will remain valid after normalization. The Python code below shows how the RK4 method has been implemented to solve the system of equations (\ref{eq:RK4}) with original values of the physical situation, the axes of the graphs have been normalized in terms of the Earth radius.\\

\begin{lstlisting}[style=Python, mathescape, caption={
RK4 Implementation in python to solve the equations of motion for a charged particle in interaction with the Earth's magnetosphere (dipole approximation)}.]
import numpy as np

#====== Def. key constants ======
RT   = 6371000.     # Earth radius [m]
m_p  = 1.67E-27     # mass of proton [kg]
m_e  = 9.109E-31    # mass of electron [kg]
qe   = 1.602E-19    # charge of proton [C]
phi  = 11.70*np.pi/180.   # Magnetic dipole tilt [rad]
th   = 23.67*np.pi/180.   # Earth's obliquity [rad]
mu   = -7.94e22*np.array([.0, np.sin(phi), np.cos(phi)
       ]) # Earth's magnetic moment [A m2]
R0dip = np.array([0.0, 0.0, 0.0]) # Dipole moment 
        location
M0=1.0E-7 #mu0/4pi
#====== Def. magnetic field at point r ======
def B(R,R0,mu):
 r = np.array([R[0]-R0[0],R[1]-R0[1],R[2]-R0[2]])*RT
 rmag = np.sqrt(r[0]**2 + r[1]**2 + r[2]**2)
 Bfield=M0*(3.0*r*np.dot(mu,r)/(rmag**5)-mu/(rmag**3))
 return Bfield

#====== setup time steps ======
dt = 0.0001
tf = 5000.
Nsteps = int(tf/dt)

#====== vector initialization ======
t  = np.zeros(Nsteps)
rp = np.zeros((len(t), 3))
vp = np.zeros((len(t), 3))

#====== Setup a charged particle ======
m = 4.0*m_p
q = 2.0*qe

#====== Define initial conditions ======
t[0]    = 0.
rp[0,:] = np.array([5. , 5., 5.])
vp[0,:] = np.array([1. , 1., 1.])

#====== RK4 implementation ======
for i in range(1, Nsteps):
  rp1 = rp[i-1,:]
  vp1 = vp[i-1,:]
  ap1 = q/m * np.cross(vp1, B(rp1,R0dip,mu))

  rp2 = rp[i-1,:] + 0.5*vp1*dt
  vp2 = vp[i-1,:] + 0.5*ap1*dt
  ap2 = q/m * np.cross(vp2, B(rp2,R0dip,mu))

  rp3 = rp[i-1,:] + 0.5*vp2*dt
  vp3 = vp[i-1,:] + 0.5*ap2*dt
  ap3 = q/m * np.cross(vp3, B(rp3,R0dip,mu))

  rp4 = rp[i-1,:] + vp3*dt
  vp4 = vp[i-1,:] + ap3*dt
  ap4 = q/m * np.cross(vp4, B(rp4,R0dip,mu))

  rp[i] = rp[i-1,:] + (dt/6.0)*(vp1+2*vp2+2*vp3+vp4)
  vp[i] = vp[i-1,:] + (dt/6.0)*(ap1+2*ap2+2*ap3+ap4)
  t[i]  = dt*i
\end{lstlisting}

\begin{figure}[h!t]
\centering\includegraphics[width=72mm]{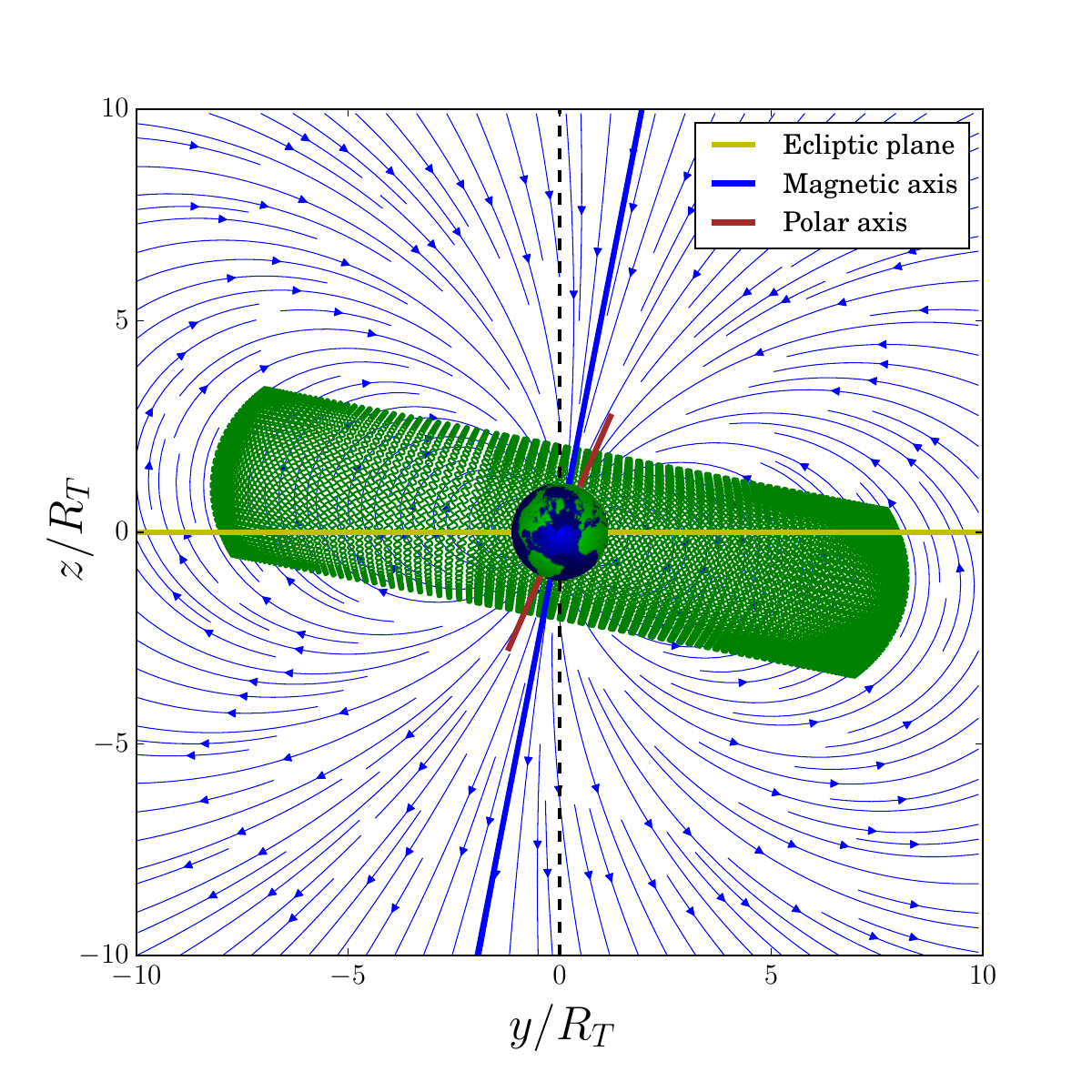}
\caption{Trajectory of an $\alpha$-particle forming a radiation belt around the Earth.}
\label{fig:Fig2Plane}
\end{figure}

\begin{figure}[h!t]
\centering\includegraphics[width=72mm]{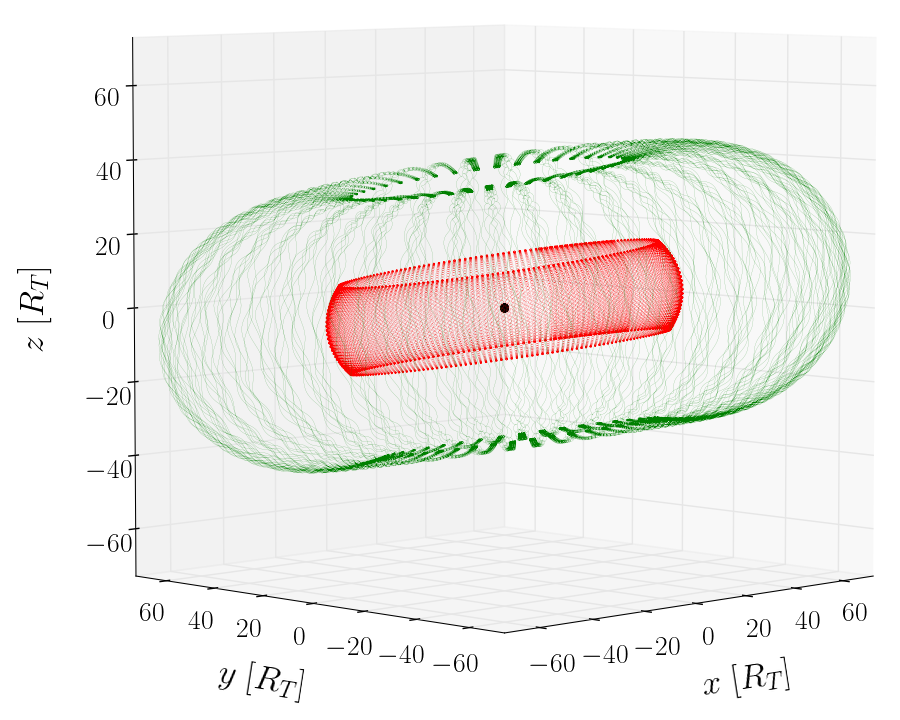}
\caption{3D trajectories of charged particles ($\alpha$ in red and $\beta$ in green) forming radiation belts around the Earth.}\label{fig:Superposition}
\end{figure}

\section{Discussion}
Table \ref{Tab:TrajectoriesParticles} shows the simulated trajectories of $\alpha$ particles and electrons immersed in the Earth's magnetosphere. A result of the simulation is a radiation ring formed by $\alpha$ particles with same orientation of the Earth's equatorial plane and symmetry axis along the vector magnetic moment as Fig. \ref{fig:Fig2Plane} shows. Depending of the mass-to-charge ratio of charged particles, the radius of the toroid is different, by instance $\alpha$ particles have orbits closer to the Earth than $\beta$ particles as illustrated in Fig. \ref{fig:Superposition}. To ensure the convergence of the numerical solution of the equations of motion, the relative difference in the total energy of the system has been plotted. This quantity was computed using $\Delta E=100|E_{teo}-E_{num}|/E_{teo}$ and in all cases it is within a uncertainty range around 1\% for the full time code execution.\\

Trajectories in configuration space gives an idea about confinement regions generated by trapped particles and also about the possible radiation rings formed around the Earth. However in some cases the behavior of physical systems can be better understood by representing the phase space in each coordinate (see Tab. \ref{Tab:TrajectoriesParticles}). In general these diagrams show information on the position and velocity of the particle allowing characterize in a unique way all possible states of the system. Each state is characterized by the position and velocity of the particle, hence it can be visualized the forbidden and permitted regions in which particles create a surrounding surface around the Earth. In addition it is possible to detect if the system dissipates energy if spiral curves appear in phase space, this is not the case in this paper since the system is conservative (there are no dissipative forces) and the magnetic field does not work and only changes the trajectory of particles.\\

On the behavior of paths in the phase space, if the movement of a point is periodic the system returns to its original state after complete a cycle. It means the representation of its trajectory in the phase space is a closed curve, moreover it is observed that the trajectory in the phase space is symmetric with respect to the vertical axis (velocity). This particular symmetry refers that movement of the particle is the same in clockwise direction as in the opposite direction, and three cases may occur: oscillations, rotations or a combination of both.\\

Considering the previous ideas it can be expected that given an initial condition there will be a single trajectory in the phase space that guides the particle. In this sense many concepts can be generalized taking into account the theory of dynamical systems and a full description from the analytical mechanics should be considered. In this case chaos may appear in some periodic trajectories, but a more elaborate mathematical tools are required to analyze these situations (such as Lyapunov coefficients, Poincaré maps among other topological structures).\\

The execution times used in the simulation are enough to describe stable trajectories with convergence rate higher than 99\%. From phase space diagrams it is very important to highlight that if two or more paths are intersect then the total energy at that particular point is not well defined (degenerate system). In this situation the system would have more than one energy value associated with a single state of the system, implying singularities in the equations of motion). The results obtained in the Tab. \ref{Tab:TrajectoriesParticles} show a high confident level of the numerical method used in the simulation and it is consistent with the physical and phenomenological explanation given in precedent sections. These results make evident that using simple approximations is possible to describe in a simple way the Van Allen belt formation through a classical model for charged particles interacting with a magnetic field.\\

\section{Conclusions}
The simulation of a particular electromagnetic phenomena, such as the movement of charged particles in the Earth's magnetosphere and its approach to the understanding of the formation of Van Allen belts, is an illustrative exercise to show how to solve a problem without analytical solution. From a pedagogical perspective this study provides support for the use of computer simulations as an educational resource important in the active/interactive learning processes, considering a qualitatively and quantitatively description of the real phenomenon. All this becomes relevant from many points of view since it allows a better understand of fundamental concepts on electromagnetism like magnetic field, Lorentz force law and the Biot-Savart law. Likewise simulations have the computational-kindness to create more complex and realistic models of physical situations still in research, helping to structure students' conceptions and encourage students involved in courses of electricity and magnetism at undergraduate level. Finally this approach can be considered as a part of a strategy to teach and learn the electromagnetic theory interactively using simulations as a pedagogical tool, focusing in the relationship model-simulation.

\section*{Acknowledgments}
We thank the PhD program in engineering and the ``Centro de Computación de Alto Desempe\~no'' (CECAD) from Universidad Distrital Francisco Jos\'e de Caldas for providing the infrastructure to develop this work. 

\bibliographystyle{unsrt}
\bibliography{biblio}

\begin{thebibliography}{10}

\bibitem{dragt1965trapped}
AJ~Dragt.
\newblock Trapped orbits in a magnetic dipole field.
\newblock {\em Reviews of Geophysics}, 3(2):255--298, 1965.

\bibitem{howard1999global}
JE~Howard, M~Hor{\'a}nyi, and GR~Stewart.
\newblock Global dynamics of charged dust particles in planetary
  magnetospheres.
\newblock {\em Physical Review Letters}, 83(20):3993, 1999.

\bibitem{inarrea2004keplerian}
Manuel I{\~n}arrea, V{\'\i}ctor Lanchares, Jes{\'u}s~F Palaci{\'a}n, Ana~I
  Pascual, J~Pablo Salas, and Patricia Yanguas.
\newblock The keplerian regime of charged particles in planetary
  magnetospheres.
\newblock {\em Physica D: Nonlinear Phenomena}, 197(3):242--268, 2004.

\bibitem{olson2006changes}
Peter Olson and Hagay Amit.
\newblock Changes in earth’s dipole.
\newblock {\em Naturwissenschaften}, 93(11):519--542, 2006.

\bibitem{stormer1907trajectoires}
Carl St{\"o}rmer.
\newblock Sur les trajectoires des corpuscules {\'e}lectris{\'e}s dans
  l'espace. applications {\`a} l'aurore bor{\'e}ale et aux perturbations
  magn{\'e}tiques.
\newblock {\em Le Radium}, 4(1):2--5, 1907.

\bibitem{stormer1934trajectoires}
C.~{St{\o}rmer}.
\newblock {On the Trajectories of Electric Particles in the Field of a Magnetic
  Dipole with Applications to the Theory of Cosmic Radiation. Third
  Communication. With 3 Figures in the Text}.
\newblock {\em Astrophysica Norvegica}, 1:1, June 1934.

\bibitem{NASA}
National~Aeronautics NASA and Space Administration.
\newblock Earth’s magnetosphere.
\newblock
  \url{http://science.nasa.gov/newhome/headlines/guntersville98/images/mag_sketch_633.jpg},
  2007 (Retrieved 20.10.2015).

\bibitem{Davis2009}
Artice~M. Davis.
\newblock The field concept in amp\'ere's magnetostatics.
\newblock {\em American Journal of Physics}, 77(8):721--729, 2009.

\bibitem{Bezerra2012}
M~Bezerra, W~J~M Kort-Kamp, M~V Cougo-Pinto, and C~Farina.
\newblock How to introduce the magnetic dipole moment.
\newblock {\em European Journal of Physics}, 33(5):1313, 2012.

\bibitem{saletan1998classical}
J~Saletan and J~Jos{\'e}.
\newblock Classical dynamics: A contemporary approach, 1998.

\bibitem{betts1998survey}
John~T Betts.
\newblock Survey of numerical methods for trajectory optimization.
\newblock {\em Journal of guidance, control, and dynamics}, 21(2):193--207,
  1998.

\bibitem{mazumdar1988analysis}
D~Mazumdar and RIL Guthrie.
\newblock An analysis of numerical methods for solving the particle trajectory
  equation.
\newblock {\em Applied mathematical modelling}, 12(4):398--402, 1988.

\bibitem{mcguire2003using}
George~C McGuire.
\newblock Using computer algebra to investigate the motion of an electric
  charge in magnetic and electric dipole fields.
\newblock {\em American Journal of Physics}, 71(8):809--812, 2003.

\bibitem{Zhaoqi2017}
Zhaoqi Zhou, Yang He, Yajuan Sun, Jian Liu, and Hong Qin.
\newblock Explicit symplectic methods for solving charged particle
  trajectories.
\newblock {\em Physics of Plasmas}, 24(5):052507, 2017.

\bibitem{ioanoviciu2015new}
Damaschin Ioanoviciu.
\newblock New method to determine proton trajectories in the equatorial plane
  of a dipole magnetic field.
\newblock {\em SpringerPlus}, 4(1):130, 2015.

\end{thebibliography}
\end{document}